\documentstyle[sprocl,psfig]{article}

\def\lsim{\mathrel{\mathpalette\vereq<}}
\def\gsim{\mathrel{\mathpalette\vereq>}}
\catcode`\@=11
\def\vereq#1#2{\lower3pt\vbox{\baselineskip1.5pt \lineskip1.5pt
\ialign{$\m@th#1\hfill##\hfil$\crcr#2\crcr\sim\crcr}}}
\catcode`\@=12

\def\Journal#1#2#3#4{{#1} {\bf #2}, #3 (#4)}

\def\NPB{{\em Nucl. Phys.} B}
\def\PLB{{\em Phys. Lett.}  B}
\def\PRL{\em Phys. Rev. Lett.}
\def\PRD{{\em Phys. Rev.} D}

\def\be{\begin{equation}}
\def\ee{\end{equation}}
\def\bea{\begin{eqnarray}}
\def\eea{\end{eqnarray}}

\begin{document}

\begin{flushright}
LBL-38891\\
UCB-PTH-96/21\\
~
\end{flushright}

\title{PHYSICS PROSPECTS \footnote{Invited plenary talk given at 3rd
International Workshop on Physics and Experiments
with $e^+ e^-$ Linear Colliders, 8-12 Sep 1995 , Iwate, Japan}\\
--- WHY DO WE WANT A LINEAR COLLIDER? --}

\author{HITOSHI MURAYAMA\footnote{This work was
supported in part by the Director, Office of  
Energy Research, Office of High Energy and Nuclear Physics, Division of 
High Energy Physics of the U.S. Department of Energy under Contract 
DE-AC03-76SF00098 and in part by the National Science Foundation under 
grant PHY-90-21139.}}

\address{Department of Physics, University of California\\
Berkeley, CA 94720, USA\\
{\rm and}\\
Theoretical Physics Group, Lawerence Berkeley Laboratory\\
University of California, Berkeley, CA 94720, USA}




\maketitle\abstracts{
The need to understand physics of electroweak symmetry breaking is
reviewed.  An electron positron linear collider will play crucial roles
in that respect.  It is discussed how the LHC and a linear collider need
each other to understand symmetry breaking mechanism unambiguously.  Two
popular scenarios, supersymmetry and technicolor-like models, are
used to demonstrate this point.}

\section{Introduction}

Now the long-awaited {\it top quark}\/ is discovered.

I have been trying to tell my non-physicists friends how significant
this result is.  Some told me back before I began explanations,
\begin{quote}
``Oh, yeah, I read the story in {\it The New York Times}.\/  I thought
particle physics is over now.''
\end{quote}

Everbody in this audience knows the impression my friend got from the
newspaper article is {\it wrong}.\/  But how wrong?  I believe it is
useful to start my talk by discussing where we are now.  Then it becomes
clearer where we are heading, and what experimental facilities we need
to achieve our goals.  

\section{Where are we?}

As everybody in this audience knows, there are (at least) two important and
exciting progress in particle physics in recent years.  The first one is
of course the discovery of the top quark.  The indication of its
existence goes back to the discovery of $\tau$-lepton in
1975\footnote{Of course, I need to mention the Kobayashi--Maskawa theory
of CP violation in year 1973 which gave us a theoretical motivation for
the third generation.} for which
Martin Perl was awarded the 1995 Nobel Prize in Physics.  The existence
of the third generation lepton, combined with the theoretical
requirement of the anomaly cancellation, implies there must exist bottom
and top quarks.  The bottom quark was discovered by a group led by Leon
Lederman in 1976.  This was the start of the long search for the top
quark.  The electron positron colliders, PEP, PETRA, and TRISTAN
established that the bottom quark has an isospin $g_A = -1/2$, which
indicated the bottom quark has (at least) one partner with charge $2/3$.
The precision electroweak measurements at LEP and SLC determined the
mass range of the top quark to be $m_t = 169^{+23}_{-27}$~GeV (PDG 94)
indirectly.  And finally, in year 1995, the long sought-after particle
was discovered by CDF and D0 experiments at Fermilab, Tevatron.  This
was a great confirmation of our current understanding in particle
physics based on SU(2)$\times$U(1) gauge theory.

\begin{figure}
\centerline{
\psfig{file=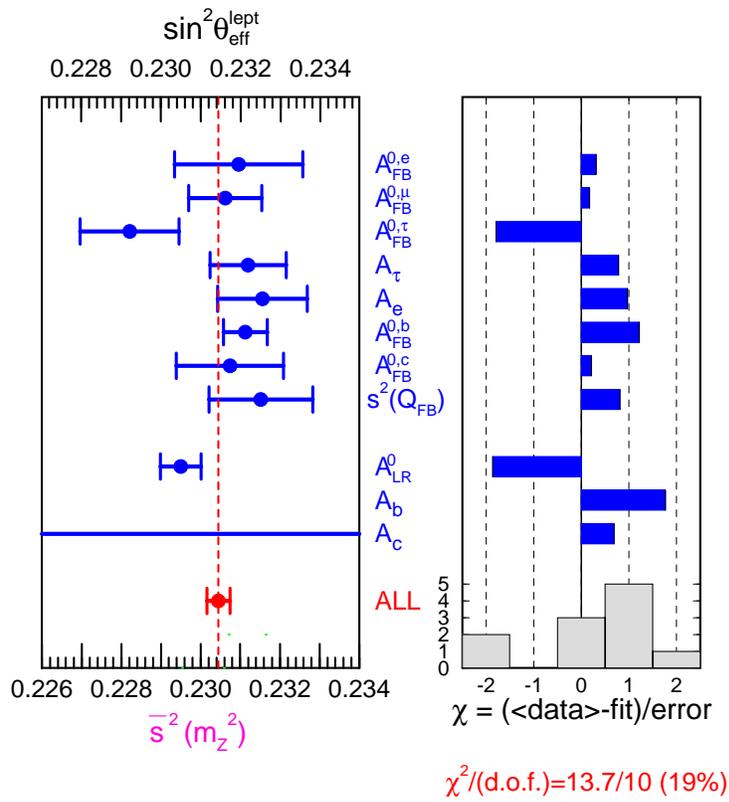,width=0.8\textwidth}
} 
\caption[Matsumoto]{A plot showing the universality of $\sin^2 \theta_W$
in many different experimental observables (asymmetries) from
experiments at LEP and SLC.  Made by S.~Matsumoto.}
\label{Matsumoto}
\end{figure}

Another important progress was that the universality of weak
interactions was established at an extremely high precision.  The ratio
of strengths of the charged current interactions (CC) of quarks and
leptons are known to be
\begin{equation}
\frac{\mbox{strengh of CC for up quark}}
{\mbox{strengh of CC for $e$ and $\mu$}} = 
|V_{ud}|^2 + |V_{us}|^2 + |V_{ub}|^2 = 0.9983 \pm 0.0015,
\end{equation}
which supports strongly a universal strengh.  More remarkably,
experiments at LEP and SLC supplied many independent measurements of the
strengh of weak neutral current, $\sin^2 \theta_W$, which are consistent
with each other with a healthy fluctuation (see Fig.~\ref{Matsumoto}).  

The universality of both charged and neutral weak interactions, combined
with the discovery of predicted top quark, strongly suggests that the
weak interactions are described by a gauge theory.  Or in other words,
the $W^\pm$ and $Z^0$ bosons are gauge bosons.  This is a natural
analogue of the fact that other known universal forces, namely gravity,
which acts univerally on all bodies (equivalence principle), and
electromagnetism, which gives universal Coulomb force which does not
depend on the properties of matter but only on their electric charges,
are described by gauge theories.  In fact, the universality was the main
motivation for Glashow to describe the weak interactions by his
SU(2)$\times$U(1) gauge theory, or electroweak theory.

However, here we encounter a contradiction.  Other known gauge forces,
gravity and electromagnetism, are known to be long-ranged.  For
instance, the range of electromagnetism is known to be larger than
1~kpc from the fact that the galactic magnetic field extends over a
distance of this order of magnitude.  On the other hand, the weak forces
are very short-ranged; they do not act beyond a distance of
$10^{-16}$~cm.  

The short-rangedness of weak interactions tells us that the electroweak
gauge symmetry has to be broken.  The ``vacuum'' is filled with a
condensate which is electrically neutral, but feels weak forces.  Since
the condensate is electrically neutral, photon can freely travel in the
``vacuum'' without knowing there exists a condensate.  On the other
hand, the carriers of weak forces, $W^\pm$ and $Z^0$ bosons, cannot
travel freely in the ``vacuum'' because their motion is disturbed by the
condenstate which feels the weak forces.
Because of this disturbance due to the condensate, $W^\pm$ and $Z^0$
bosons cannot travel far, and the weak interactions become short-ranged.

In the Standard Model, the condensate is assumed to be a spinless boson
which acquires a vacuum expectation value.  In order to generate this
condensate, one introduces a potential for the spinless boson and
assumes it has a double-well form such that the minimum of the potential
lies where the boson has a non-vanishing value.  However, this
``explanation'' leaves many questions open.  First of all, why such a
spinless boson exists, while we have not seen any elementary spinless
bosons in nature yet.  Even if we accept the existence of such a
spinless boson, it is mysterious why it has such a special form of the
potential which is {\it designed}\/ to generate a non-vanishing vacuum
expectation value.  Furthermore, we know that the masses of elementary
fermions, leptons and quarks, vary between almost six orders of
magnitudes.  The ``explanation'' of this diversity in the Standard Model
is that the top quark, the heavest particle, interacts with the
condenstate strongly and its motion is substantially disturbed as the
$W^\pm$ and $Z^0$ bosons, while the electron, the lightest charged
particle, interacts only very weekly with the condensate so that it does
little harm to the motion of electrons.  It is left completely
unexplained why the quarks and leptons interact with the condensate with
so different strengths.

Because of this unsatisfactory nature, the Standard Model cannot be the
whole story of nature.  A true theory of the electroweak symmetry
breaking, the mechanism which makes the weak interactions short-ranged,
must explain why it is broken.  I believe this is an {\it experimental
question}\/ which has to be answered by studing the properties of
$W^\pm$, $Z^0$ bosons and search for new phenomena below TeV scale.
Despite the efforts by both our experimental and theoretical colleagues
for more than two decades, we have little clue on this point.  The
next generation experiments have to planned so that they will be able to
give us clues to answer this question.

Now what kinds of experimental facilities are needed to explore the
physics of electroweak symmetry breaking?  To discuss this point, I use
two popular scenarios, supersymmetry and technicolor, as
representative models.  Even though we probably have not exhausted all
theoretical possibilities to explain the electroweak symmetry breaking,
an experiment cannot cover a scenario which we could not think of so
far if it cannot cover already-known scenarios.  Therefore, future
facilities have to be designed to cover {\it at least}\/ these two
scenarios of electroweak symmetry breaking.

In Table~1, I listed ``explanations'' to various questions in the
Standard Model in both scenarios.  As one can easily see, both of them
predict very rich phenomena at TeV scale.  Moreover, both of them leave
further fundamental questions to physics at yet higher energies which
are very distinct.  Therefore, we will obtain very useful clues to the
physics much above TeV scale once we understand the physics of
electroweak symmetry breaking.  TeV scale machines will give us hints on
physics at much higher energy scales.  

\begin{table}
\caption[scenarios]{``Explanations'' given to basic questions on physics of
electroweak symmetry breaking in various theory scenarios.  The contents
in this table are meant to be examples, rather than representative ones.
Especially those on fermion masses are controversial.}
\begin{center}
\scriptsize
\begin{tabular}{|cccc|}
\hline
& Standard Model & Supersymmetry & Technicolor\\
\hline \hline
Existence 
& Only scalar boson 
& Just one of many 
& No Higgs boson.  
\\
of Higgs 
& introduced just to 
& scalar bosons, 
& There are only fermi-
\\
boson 
& break EW symmetry
& nothing special
& ons and gauge bosons.
\\ \hline
Why electro-
& by an 
& $m^2$ driven negative
& new strong 
\\
weak symmetry
& ad hoc choice 
& dynamically by top
& force binds fermions 
\\
is broken
& $m^2 < 0$
& quark Yukawa coupling
& to let them condense
\\ \hline
pattern of 
& choose size of 
& sequential breaking of
& further new 
\\
quark, lepton 
& Yukawa couplings 
& flavor symmetry just
& forces at 1 to 1000 
\\
masses
& to reproduce them
& below the Planck scale
& TeV scales
\\ \hline
new 
& 
& superpartners of 
& resonances at 1--10
\\
phenomena
& only a Higgs boson
& all known particles 
& TeV, PNGBs and new
\\
& 
& below TeV scale
& fermions at 0.1--1 TeV
\\
\hline
\end{tabular}
\end{center}
\end{table}

As clear from the Table~1, physics of electroweak symmetry breaking
is necessarily rich and complex.  The challenge in designing the next
generation experiments is to disentangle such complex signatures.  In
later sections, I discuss the case of supersymmetry scenario
and ``techicolor-like'' scenario to see how well we can understand
physics of electroweak symmetry breaking at the LHC and a possible
future electron positron linear collider.  It will be argued that both
types of colliders are necessary to understand rich physics of
electroweak symmetry breaking unambiguously; they play different roles,
and work together leading us to decide yet-further future direction of
the field.

\section{Light Higgs and supersymmetry case}

Let me take the Minimal Supersymmetric Standard Model (MSSM) as an example
below.  There are five Higgs bosons in this model,
\begin{displaymath}
h^0, H^0, A^0, H^+, H^-,
\end{displaymath}
and the mass of the lightest neutral scalar $h^0$ has to be smaller than
$m_{h^0} \lsim 130$~GeV including radiative corrections.\cite{Okada} It
decays primarily into $b\bar{b}$, and into $\gamma\gamma$ with a
branching fraction of $\sim 10^{-3}$ or less.

\begin{figure}
\centerline{
\psfig{file=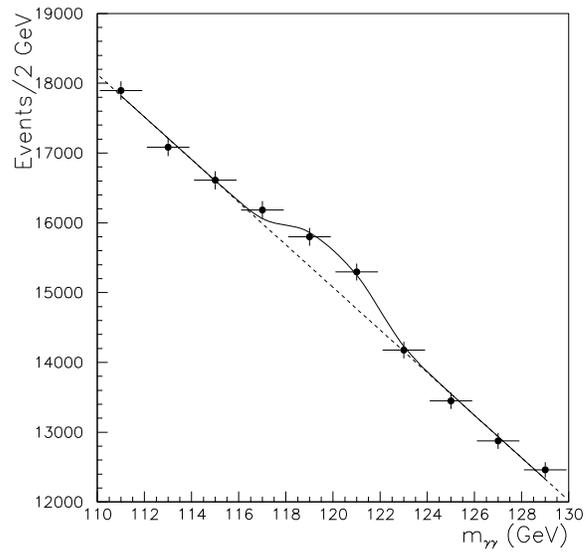,width=0.7\textwidth}
}
\caption[FPsigbck]{Expected $h\rightarrow\gamma\gamma$ signal for the
Standard Model Higgs $m_h = 120$~GeV, combined with the prompt
$\gamma\gamma$ background, assuming an integrated luminosity of
$10^5$~pb$^{-1}$.  Taken from ATLAS TDR.\cite{ATLAS-TDR}}
\label{gamma-gamma}
\end{figure}

The LHC will see the signal of a light neutral scalar decaying into
$\gamma\gamma$ with an impressive capability even in the high luminosity
environment (Fig.~\ref{gamma-gamma}).  The ATLAS and CMS experiments
will discover the Standard Model Higgs boson over the entire mass range
above LEP2 reach up to 600~GeV or so.  The $\gamma\gamma$ rate is in
general lower in the Minimal Supersymmetric Standard Model than in the
Standard Model, but still they will cover most of the parameter space.
This is a highly significant capability of these experiments.

\begin{figure}
\centerline{\psfig{file=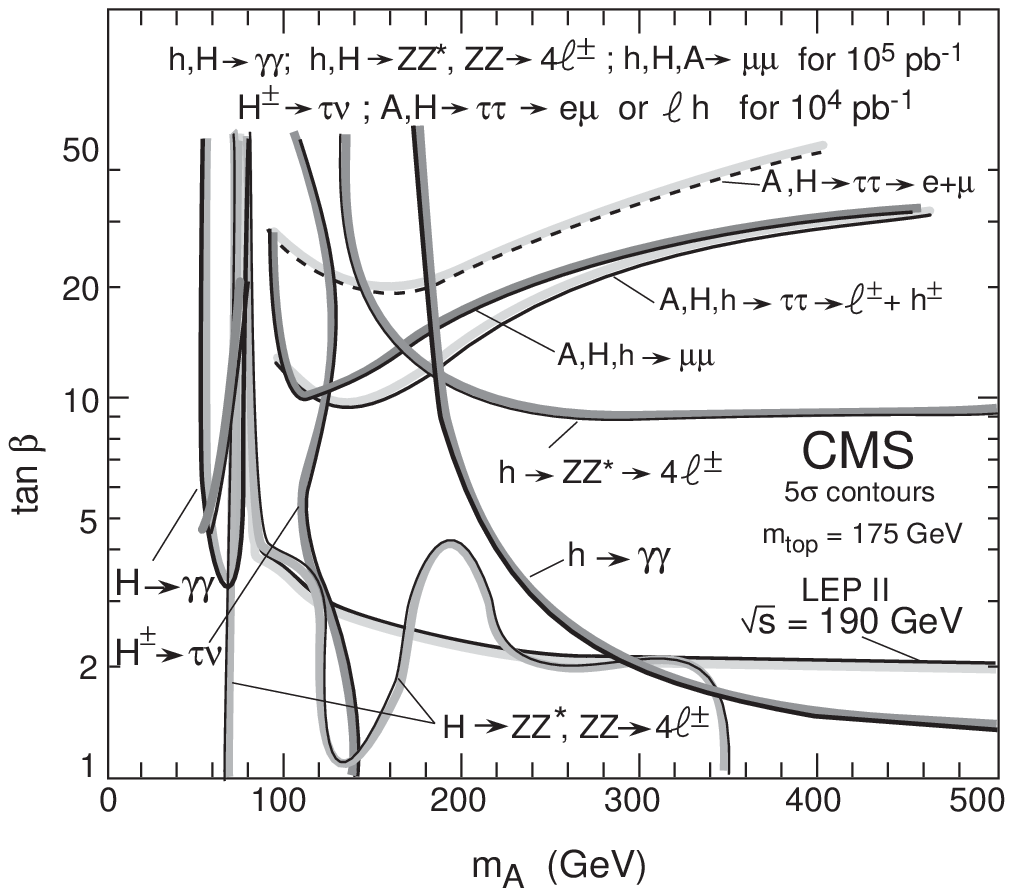,width=0.7\textwidth}}
\caption[CMS]{The coverage of the MSSM Higgs parameter space by
CMS experiment.\cite{CMS-TDR}}
\label{CMS}
\end{figure}

However I still have a worry if there were only LHC and no electron
positron collider.  It is not the fact that there remains a hole in the
MSSM parameter space (Fig.~\ref{CMS}), as some people
emphasize.  This may be filled by running experiments for 3
years at high luminosity and combine two experiments.\cite{nohole}  My 
worry is it is not clear what we will learn either by seeing this signal
or by not seeing it.  

Suppose we will see a peak in $\gamma\gamma$ invariant mass
distribution.  I worry that it may not decide what is responsible for
electroweak symmetry breaking.  Let me first present a toy example of a
model which has nothing to do with electroweak symmetry breaking but
gives exactly the same signature and rate.  This model has a new quark
$U_L$ and $U_R$ with the same SU(3)$_C \times$ SU(2)$_L \times$ U(1)$_Y$
quantum numbers as the right-handed up quark, ({\bf 3}, {\bf 1},
$\frac{2}{3}$), and a scalar field $\phi$ which is singlet under the
standard model gauge group.  There is a Yukawa interaction between $U$
and $\phi$,
\begin{equation}
{\cal L} = \bar{U} U \phi,
\end{equation}
and a vacuum expectation value $\langle \phi \rangle \neq 0$ generates a
mass for the $U$-quark.\footnote{The absence of an explicit mass term
is natural since one can assign a $Z_2$ symmetry $\phi \rightarrow -
\phi$, $U_L \rightarrow - U_L$, and $U_R \rightarrow U_R$.}  Since $\phi$
is singlet under the standard 
model gauge group, its condensate does not give masses to $W^\pm$ and
$Z^0$, and has nothing to do with the electroweak symmetry breaking.
The production cross section of $\phi$ from $gg$ fusion via $U$-quark is
the same as that of the Standard Model Higgs boson via top quark loop
because they are the same triangle diagram; it is known that
the triangle diagram does not depend on the mass of internal fermion if
the mass of the scalar particle is less than twice the mass of fermion.  
On the other hand, $\phi$ decays mainly back to $gg$, but decays also
into $\gamma\gamma$ with a branching fraction of
\begin{equation}
\mbox{Br}(\phi \rightarrow \gamma\gamma)
	= \left( \frac{(2/3)^2 \alpha}{(4/3) \alpha_s} \right)^2
	\simeq 10^{-3},
\end{equation}
which is again the same as the Standard Model Higgs boson.  Therefore,
it remains not clear whether what we have seen is the Higgs boson or
something else.

Of course the above toy model is not a well-motivated theory.  But there
are presumably many other examples which lead to similar experimental
signatures.  For instance, one of the pseudo--Nambu--Goldstone bosons
(PNGB) in technicolor models, or techni-pions, can couple to $gg$ and
$\gamma\gamma$, and can mimic the signal.  In this case the
$\gamma\gamma$ peak does see the physics of electroweak symmetry
breaking, but its interpretation is ambiguous.  It may not establish
whether the physics is supersymmetry-like or technicolor-like.

The source of the ambiguity is that the $\gamma\gamma$ signature, or
other possible ones like $t\bar{t} h$, $b\bar{b} h$, do not test the
crucial characteristics of {\it the}\/ Higgs boson.  What are its crucial
characteristics?  There are three of them.  (1) It has to be a scalar
particle.  (2) It has a condensate in the vacuum.  (3) It generates
$m_W$ and $m_Z$.  Can we test these characteristics experimentally?

\begin{figure}
\centerline{\psfig{file=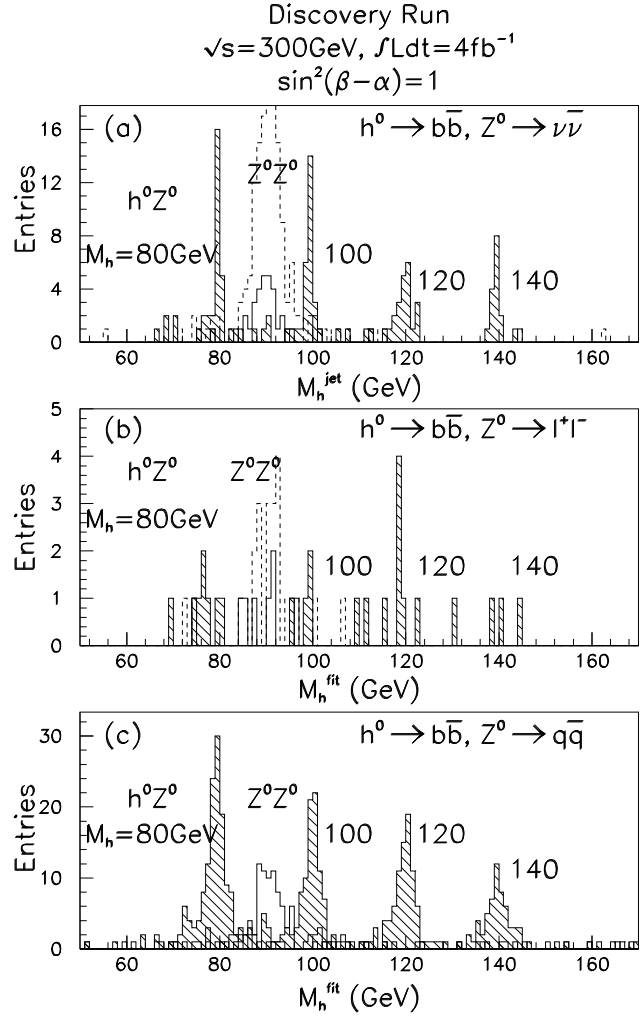,width=0.7\textwidth}}
\caption[mh]{The invariant mass distributions in $Zh$ events for the
standard model Higgs boson.\cite{JLC-I}  All decay modes of $Z$, (a)
$Z\rightarrow
\nu\bar{\nu}$, (b) $l^+ l^-$, (c) $q\bar{q}$ can be used.  The
integrated luminosity 4~fb$^{-1}$ corresponds to about a week with the
design luminosity.}
\label{JLC-I}
\end{figure}

It is not difficult to test the crucial characteristics of {\it the}\/
Higgs boson at an electron positron linear collider once it is found
(Fig.~\ref{JLC-I}). 
The most promising production process for a light Higgs boson which we
are discussing here is $e^+ e^- \rightarrow Z h$.  First of all, the
polarization asymmetry of Higgs boson production is rather small,
proportional to $1-4\sin^2\theta_W$.  The smallness tells us that there
is no significant $\gamma$-$Z$ interference, whose relative sign is
roughly opposite for different electron polarization.  Therefore we
learn that either $\gamma$ or $Z$ dominates in the process.  A small but
finite asymmetry then confirms it is $Z$-dominated, and hence the
production is due to $ZZh$ coupling.  One can check that the
final $Z$-boson is mainly 
longitudially polarized by reconstrucing $Z$ decay distribution, and hence
the $Z$-boson can be regarded as a scalar boson.  The
angular distribution of the Higgs boson is $\sin^2 \theta$ in the high
energy limit,\footnote{The distribution is $d\sigma/d\cos\theta \propto
2 + \gamma_Z^2 \beta_Z^2 \sin^2 \theta$.}  which tells
us the discovered particle is a scalar, CP-even particle.\footnote{If
the scalar particle were CP-odd, it should be produced via a
dimension-five interaction $\epsilon^{\mu\nu\rho\sigma} Z_{\mu\nu}
Z_{\rho\sigma} \phi$, and both the angular distribution of the Higgs
boson and the decay angle distribution of $Z$-boson are different.} 
Combining these observations, it establishes that the production 
occurs via $Z_\mu Z^\mu h$ operator.   Since usual scalar fields without a
condensate have only $Z_\mu Z^\mu \phi^* \phi$ coupling but no $Z_\mu
Z^\mu \phi$ coupling, the
existence of $Z_\mu Z^\mu h$ coupling implies $h$ has a vacuum
expectation value.  Finally the total production rate independent 
from the decay modes can be measured using leptonic decay of $Z$, which
gives us 4\% level measurement of the $ZZh$ coupling with 50~fb$^{-1}$
integrated luminosity.\cite{Janot-Hawaii}  If the observed particle is
{\it the}\/ Higgs boson, the $ZZh$ coupling has to be $g_Z m_Z = e
m_Z/\sin \theta_W \cos \theta_W$. 
Having $ZZh$ coupling with the right strength establishes that it
is responsible for generating $m_Z$.  In this way, one can unambiguously
establish that the observed new particle is {\it the}\/ Higgs boson.  If
the coupling is less, it contributes to a part of the $Z$ mass, and
there should be more Higgs boson(s) to generate the entire $Z$ mass.  

Furthermore, one can even measure relative branching ratios of the Higgs
boson.  Table~2 shows the expected accuracy of branching ratio
measurements with 50~fb$^{-1}$ without using polarization.  

\begin{table}
\caption[Hildreth]{The errors in branching fraction
measurement,\cite{Hildreth-Hawaii} calculated 
assuming Standard Model coupling for the Higgs boson and 50~fb$^{-1}$ of
integrated luminosity at $\sqrt{s}=400$~GeV. 
}
\begin{center}
\begin{tabular}{|c|cc|}
\hline
& $m_h = 140$~GeV & $m_h = 120$~GeV\\
Branching Fraction & Expected error & Extrapolated error\\
\hline \hline
$h \rightarrow b\bar{b}$ & $\pm 12$~\% & $\pm 7$~\% \\
$h \rightarrow WW^*$ & $\pm 24$~\% & $\pm 48$~\% \\
$h \rightarrow c\bar{c}+gg$ & $\pm 116$~\% & $\pm 39$~\%\\
$h \rightarrow \tau^+ \tau^-$ & $\pm 22$~\% & $\pm 14$~\%\\
\hline
\end{tabular}
\end{center}
\end{table}
Nakamura\cite{Nakamura-Iwate} discussed much better measurement of
$c\bar{c} + gg$ branching fraction with an aid of electron beam
polarization which can suppress the $WW$ background substantially by
employing right-handed polarization of electron.  Such a measurement may
hint that the Higgs boson is not that of the Standard Model but rather
of an extended version like the Minimal Supersymmetric Standard Model.

A truly interesting strategy is to use information from all possible
experiments, $pp$, $e^+ e^-$ and $\gamma\gamma$ colliders.  The LHC measures
the product $\Gamma(h \rightarrow gg) \Gamma(h \rightarrow
\gamma\gamma)/\Gamma_h$, while a $\gamma\gamma$ collider measures
$\Gamma(h \rightarrow \gamma\gamma)^2/\Gamma_h$.  An $e^+ e^-$ linear
collider will give us $\Gamma_h$ indirectly, knowing the $ZZh$ vertex
from the total production rate and the relative branching fraction into
$WW^*$.  Combination of all three experiments will give us
model-independent determination of $gg$ and $\gamma\gamma$ partial
widths.\cite{DPF-Higgs}  Such a measurement is of a great interest
since {\it any}\/ charged or colored particles which obtain their masses
from electroweak 
symmetry breaking contribute to these partial widths and do not decouple
even when they are heavy.  Therefore a determination of these widths may
signal the existence of heavy particles.  This is a wonderful example
how different colliders cooperate to give us useful information on
physics of electroweak symmetry breaking and {\it beyond}.\/

\section{Supersymmetry}

The search for and study of superparticles offer us the best example
where the LHC and an $e^+ e^-$ linear collider play different roles,
which combine to give us a coherent picture of physics at yet deeper
level. 

\begin{figure}
\centerline{
\psfig{file=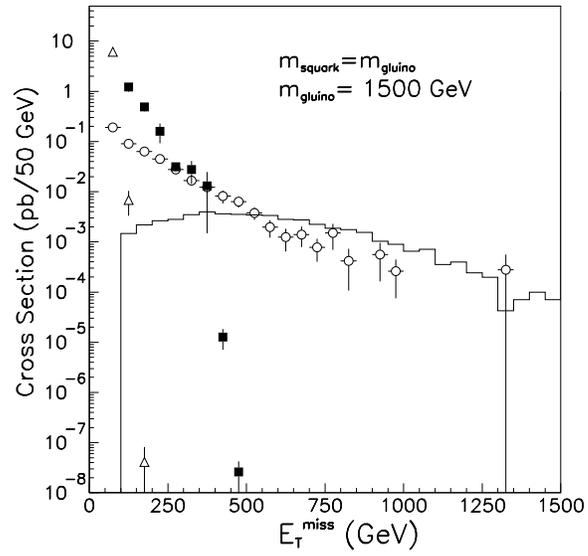,width=0.7\textwidth}
}
\caption[missET]{Missing $E_T$ distribution of the supersymmetry signal
with $m_{\tilde{q}} = m_{\tilde{g}} = 1.5$~TeV (full line) together with
physics background (open circles), instrumental background in the
pessimistic case (black squares) and the realistic case (open triangles).
Taken from ATLAS TDR.\cite{ATLAS-TDR}}
\label{missET}
\end{figure}

Discovery of supersymmetry at the LHC is regarded as a relatively easy
job.  In the ordinary framework of supersymmetry,\footnote{It is assumed
that the $R$-parity is exact and the lightest superparticle is a stable
neutralino.} the dominant signature of supersymmetry is large missing
$E_T$ with many jets (Fig.~\ref{missET}).  For instance, the gluon
fusion produces a pair of gluinos, gluinos decays into a squark and a
quark, the squark decays into a chargino and a quark, the chargino
decays into the lightest neutralino and $W$, and $W$ into jets or a
lepton and a neutrino.  Since the lightest neutralino and the neutrino
escape detectors, one sees large missing $E_T$ with many jets (and
leptons).

Similar to the case of the Higgs boson, again the interpretation of the
signal is not clear.  If one sees, in addition to the missing $E_T$
signal, like-sign dileptons, it is consistent with the ``Majorana-ness''
of gluino and the interpretation becomes more solid.  But still, it is
much more favorable if one can directly see that (1) there are new
particles with the same quantum numbers as those of known particles, (2)
their spin differ by 1/2, and (3) their interactions respect relations
required by supersymmetry.  All three are possible at an $e^+ e^-$ linear
collider in principle. 

\begin{figure}
\centerline{
\psfig{file=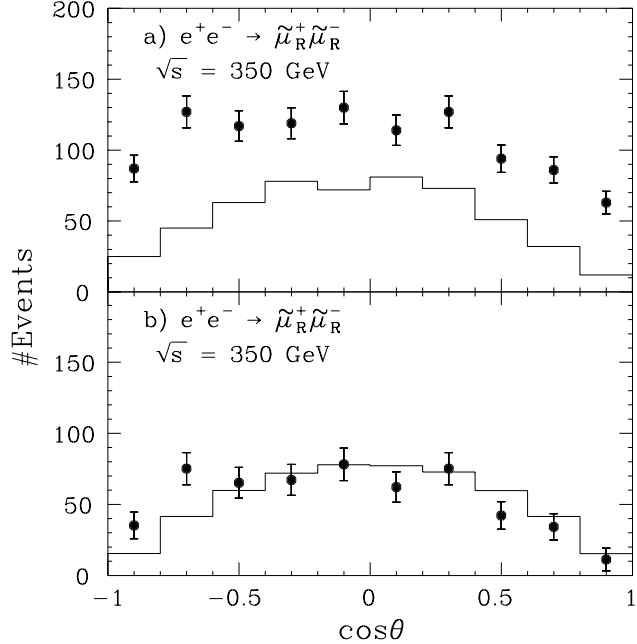,width=0.7\textwidth,angle=90}
}
\caption[smu-angle]{The angular distribution of $\tilde{\mu}_R^-$
   in $e^+e^- \rightarrow \tilde{\mu}^+_R\tilde{\mu}^-_R$
   reconstructed from final-state $\mu^\pm$ four-momenta, knowing the
   $\tilde{\mu}_R$ and $\tilde{\chi}_1^0$ masses:
   (a) with the two solutions unresolved and
   (b) with the ``background'' due to the wrong solutions subtracted.
   The histogram in (a) is the distribution of the right solutions,
   while that in (b) is the distribution of the original sample
   before selection cuts.\cite{Tsukamoto}
}
\label{smu-angle}
\end{figure}

\begin{figure}
\psfig{file=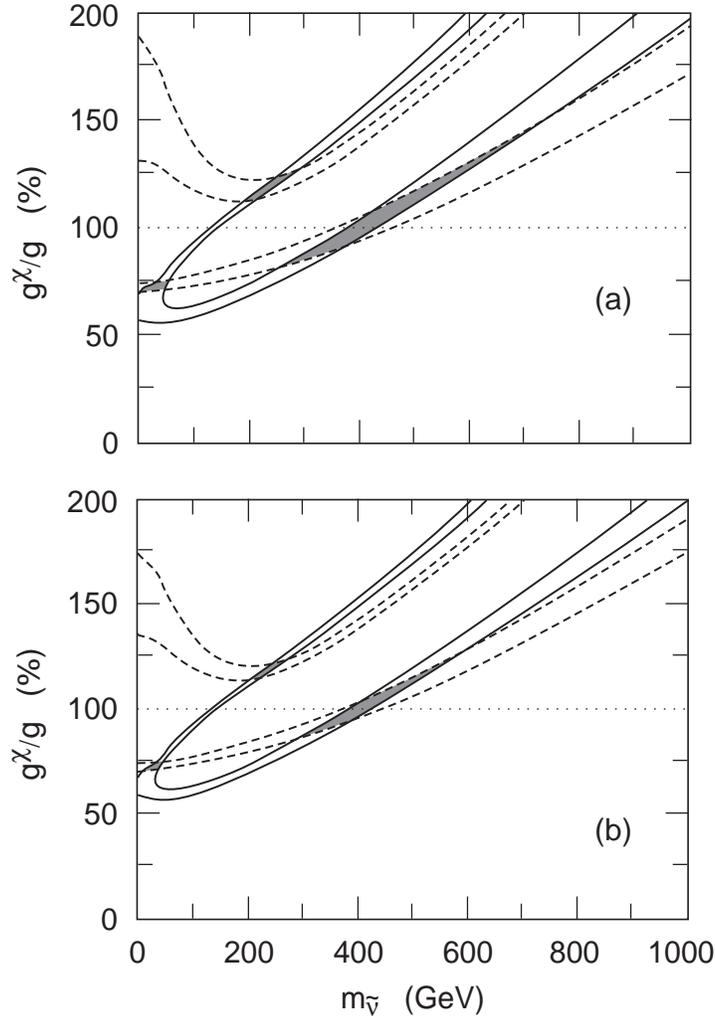,width=0.8\textwidth}
\caption[jonathan]{Allowed regions (shaded) of the $(m_{\tilde{\nu}},
g^\chi)$ plane for ${\cal L}_L$ 
= (a) 30~fb$^{-1}$ and (b) 100~fb$^{-1}$. Solid (dashed) curves are contours
of constant $\sigma_L$ ($A_{FB}^{\chi}$) that bound the allowed regions.
On the dotted lines, the SUSY relation $g^\chi = g$ is
satisfied.\cite{FPMT}
} 
\label{jonathan}
\end{figure}

Let us suppose we see sleptons at a future $e^+ e^-$ linear collider.
It is easy to determine that the sleptons have the same quantum numbers
as the leptons, just by counting the number of events.  For instance the
production of $\tilde{\mu}$ is due to $s$-channel $\gamma$, $Z$
exchange.  The total production cross section and the left-right
asymmetry completely determines the coupling of $\tilde{\mu}$ to
$\gamma$ and $Z$.  Even though $\tilde{\mu}$ decays into $\mu$ and the
lightest neutralino which escapes detection, the angular distribution of
the $\tilde{\mu}$ can be also reconstructed up to a two-fold ambiguity.
Fortunately, the ``wrong'' solution has a flat distribution which can be
subtracted.  Then one clearly sees $\sin^2 \theta$ distribution which
shows that $\tilde{\mu}$ is a scalar particle (Fig.~\ref{smu-angle}).
The goal (3) is more difficult to achieve.  Fig.~\ref{jonathan} shows a
result of a case study how well one can establish the equality between
two different couplings, the usual SU(2) gauge coupling $e$-$\nu$-$W$
and its supersymmetric version, $e$-$\tilde{\nu}$-$\tilde{W}$.  We label
the former by $g$ and the latter by $g^\chi$.  Since two couplings are
related by supersymmetry, $g=g^\chi$.  The figure shows how well one can
determine the ratio $g^\chi/g$ experimentally from a pair production of
$\tilde{W}$-like chargino.  Using the total cross section and the
forward-backward asymmetry, one obtains three regions on
$(m_{\tilde{\nu}},\,g^\chi/g)$ plane.  By combining further with
(negative) experimental search for $\tilde{\nu}$, one can select the
solution $m_{\tilde{\nu}} = 400$~GeV and $g^\chi/g=1$ consistent with
the inputs in the analysis.  In this way, one can unambiguously
establish that the new phenomenon observed is indeed due to
supersymmetry.

There is even more excitement after the discovery.  Measurement of
superparticle masses will tell us physics at very high scales, like GUT-
or Planck-scales.  The best example is the following test of grand
unified theories (GUT) using the masses of gauginos.  It is now
well-known that the measured value of $\sin^2 \theta_W$ is remarkably
consistent with the prediction of supersymmetric GUTs.  The reason why
we can test a theory at a very high scale in this case is because GUTs
predict that the three gauge coupling constants are the same, $\alpha_1 =
\alpha_2 = \alpha_3$ at the GUT-scale.  We can extrapolate the measured
gauge coupling constants to higher energies, and test whether they meet
at a single point.  Some people take this seriously, others think it
is just a numerical accident.  Now supersymmetric GUTs predict further
relations.  The masses of three gauginos, $M_1$, $M_2$, $M_3$ of bino,
wino, and gluino, respectively, also have
to be the same at the GUT-scale.  Once superparticles are discovered, we
can measure their masses, and extrapolate the measured values to higher
energies.  Then we can see whether they meet at a point.  This gives us
an independent test of GUTs from that using gauge couping constants, and
if verified, it can hardly be an accident.  

\begin{figure}
\centerline{
\psfig{file=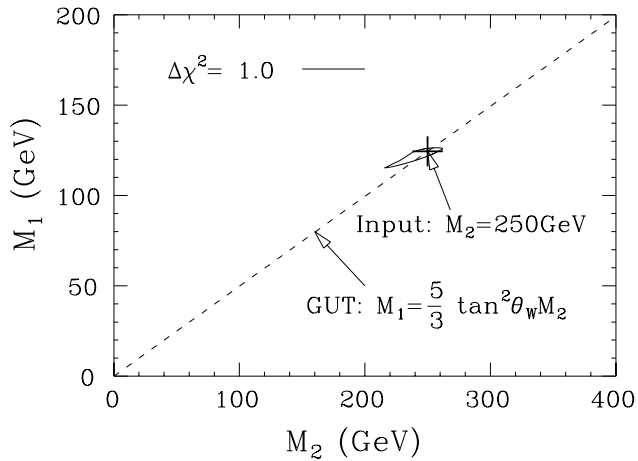,width=0.7\textwidth,angle=90}
}
\caption[guttest]{   The $\Delta \chi^2 = 1$ contour in the $M_1$-$M_2$ plane
   obtained from the global fit.\cite{Tsukamoto}
   The dotted line indicates the GUT prediction:
   $M_1 = (5/3) \tan^2 \theta_W~M_2$.
}
\label{guttest}
\end{figure}

For such a measurement of gaugino masses, an $e^+ e^-$ linear collider
with polarized electron beam is crucial.  Since gauginos $\tilde{B}$ and
$\tilde{W}$ mix with higgsinos to form two chargino and four neutralino
mass eigenstates, one needs to disentangle the mixing to measure the
masses of gauginos.  In our paper,\cite{Tsukamoto} four experimental
observables, 
$m(\tilde{\chi}_1^0)$, $m(\tilde{\chi}_1^\pm)$, $\sigma(e_R^- e^+
\rightarrow \tilde{e}_R^- \tilde{e}_R^+)$ and $\sigma(e_R^- e^+
\rightarrow \tilde{\chi}_1^- \tilde{\chi}_1^+)$ were used to extract
four parameters $M_1$, $M_2$, $\mu$ and $\tan \beta$.  The
Fig.~\ref{guttest} shows 
the accuracy how well one can extract $M_1$ and $M_2$ consistent with
the inputs, which satisfy a simple relation from GUTs.  Therefore, one
can make an important test of grand unified theories by measuring cross
sections and masses of superparticles.

Finally, the spectrum of scalar particles will tell us what kind of GUTs
it is,\cite{KMY1} or the energy scale where superymmetry is broken
(``messenger scale'').\cite{Peskin-Iwate}  Hopefully a combination of
the LHC and a linear collider would do this job.  There are several
differences between supersymmetry studies at the LHC and at a linear
collider.  The LHC produces superparticles top down.  It produces the
colored particles like gluino and squarks which are tyipcally 3 to 4
times heavier than their colorless counter parts, and they decay into
the lightest superparticle in long chains of cascades.  The decay
pattern is a complicated function of all low-lying supersymmetry
spectrum.  Therefore, signature of supersymmetry at the LHC has many
important information in it, but it is difficult to sort it out by
itself because of very complex cascades.  On the other hand, an electron
positron linear collider would produce superparticles bottom up.  As one
raises the center-of-mass energy, the lightest one will be found and
subsequently to the heavier ones.  At each stage, one studies the
newly-found superparticles in detail and determine all the parameters.
Then there is no ambiguity in studying the next superparticle because
you already know the spectrum below it.  This approach is very useful
for the colorless superparticles which are supposed to be rather light
and likely to be within the reach of a linear collider.  The LHC reach
of gluino mass up to 2~TeV roughly equals with a 1 TeV linear collider
which can find $\tilde{W}$ up to 500~GeV.  By determining basic
supersymmetry parameters at an $e^+ e^-$ and analyzing the top-down data
from the LHC using the inputs from a linear collider, we can eventually
sort out the whole superparticle spectrum.  This is a challenging, but
a very fruitful and exciting program.  And having both types of machines
is crucial in this grand program.

\section{Strong electroweak symmetry breaking sector}

Now we come to the discussion of other type of scenario, where the
electroweak symmetry is broken by a new strong force.  A representative
model is technicolor, where this new gauge interaction attracts pairs of
technifermions very strongly with each other and let them condense,
$\langle \bar{f} f \rangle \neq 0$.  The generic signatures of this
scenario are: (1) no light Higgs boson (below, say, $600$~GeV),
(2) the scattering between two longitudinal $W$ bosons become strong at
higher energies (say, $E \gsim 1.8$~TeV), and (3) there possibly are
resonances due to new strong interactions (techni-$\rho$ decaying into
$W^+ W^-$ or $W^\pm Z^0$, techni-$\omega$ into $Z^0 \gamma$, etc).  

First general statement on this scenario is that all experimental
signatures are rather rare and weak, and it will be difficult for
experiments to see the effects of new strong interaction.
Table~\ref{W+W+} shows 
the expected event rates for several different models along with the
size of the Standard Model background.  Even though it is likely that
one can see certain excess in like-sign dilepton with large missing
$E_T$, it may not be easy to directly interpret it as a signal of strong
$WW$ interaction.

\begin{table}
\caption[W+W+]{Expected numbers of events of like-sign dilepton after cuts in
the $W^+_L W^+_L$ search for an integrated luminosity of
$10^5$~pb$^{-1}$ and for different models, taken from ATLAS 
TDR.\cite{ATLAS-TDR} 
The forward jet tag is used.}
\begin{center}
\begin{tabular}{|c|c|}
\hline
Model & Number of events\\
\hline
\hline
Standard Model ($m_h = 1$~TeV) & 23 \\
\hline
Rescaled $\pi^+ \pi^-$ scattering & 25 \\
\hline
Low energy theorm (LET) & 39 \\
\hline
Sharp-cutoff unitarization & 40 \\
\hline
O(2N) Higgs-Goldstone model & 15 \\
\hline
Standard Model background & $\lsim 46$ \\
\hline
\end{tabular}
\end{center}
\label{W+W+}
\end{table}

If one also has an electron positron linear collider in addition to the
LHC, this difficult signal becomes convincing.  A linear collider can
unambiguously prove the absense of any kinds of light Higgs boson below
its kinematic reach, $m_h \lsim 0.9 (\sqrt{s} - m_Z)$.  The {\it
absense}\/ of a Higgs boson, combined with an excess in like-sign
dilepton, implies a strongly interacting electroweak symmetry breaking
sector.  Recall that it is not easy to establish the absence of a light
Higgs boson at the LHC alone.  There is a small hole in the MSSM
parameter space which is not easy to cover (Fig.~\ref{CMS}).  Also, the
Higgs boson may 
decay mainly invisibly, which reduces the $\gamma\gamma$ signature
substantially.  The invisible decay is not specific to the
supersymmetric models, where Higgs bosons may decay into a pair of
neutralinos, but also possible in other models as well.  For instance if
the fourth generation exists with little mixing to lower generations,
and if $2 m_{\nu_4} < m_H < 2 m_{l_4} < 2 m_{q_4}$, the Higgs boson
decays mainly into $\bar{\nu}_4 \nu_4$ and is hard to be detected.  One
can also look for associate production processes like $t\bar{t}H$, $WH$
even in this case;\cite{invisible} but it seems to be not easy to
convince ourselves there is {\it no}\/ Higgs boson.  On the other hand,
an invisibly decaying Higgs boson can be easily seen at a linear
collider using $Zh$ production with $Z$ decaying
leptonically.\cite{Janot-Hawaii} 

So far the role a linear collider plays may seem secondary, just to give
a supportive evidence by proving there is no light Higgs boson.  But
there are more active roles an electron positron linear collider can
play as well.  

Table~\ref{eeWW} shows the significance of strong $WW$ scattering
studied at 
an $e^\pm e^-$ collider with $\sqrt{s} = 1.5$~TeV and an integrated
luminosity of 200~fb$^{-1}$.  The statistical significance is comparable
or sometimes better than that at the LHC.

\begin{table}
\caption[eeWW]{Total numbers of $W^+W^-, ZZ \rightarrow  4$-jet
signal $S$ and background $B$ events calculated for  a 1.5~TeV
$e^\pm e^-$ linear collider with  integrated luminosity 200~fb$^{-1}$
after cuts. The 
statistical significance $S/\sqrt B$ is also given.
The hadronic branching fractions of $WW$ decays and the $W^\pm/Z$
identification/misidentification are included.
S/N is Improved by using $100\%$ polarized $e^-_L$ beams in a 1.5~TeV
$e^+e^-/e^-e^-$ collider.\cite{BCHP}}
\begin{center}
\small
\begin{tabular}{|l|c|c|c|c|}
\hline
channels & SM  & Scalar & Vector   & LET  \\
& $m_H=1$ TeV & $M_S=1$ TeV & $M_V=1$ TeV &\\
\hline
$S(e^+ e^- \to \bar \nu \nu W^+ W^-)$
& 330   & 320   & 92  & 62  \\
$B$(backgrounds)
& 280    & 280   & 7.1  & 280  \\
$S/\sqrt B$ & 20 & 20 & 35 & 3.7 \\
\hline
$S(e^+ e^- \to \bar\nu \nu ZZ)$
&  240  & 260  & 72  & 90   \\
$B$(backgrounds)
& 110    & 110   & 110  & 110  \\
$S/\sqrt B$ & 23 & 25& 6.8& 8.5\\
\hline
\hline
$S(e^- e^-_L \to \nu \nu W^- W^-)$  & 54 & 70 & 72 & 84 \\
$B$(background) & 400 & 400 & 400 & 400\\
$S/\sqrt B$ & 2.7 & 3.5 & 3.6 & 4.2 \\
\hline
$S(e^-_L e^-_L \to \nu \nu W^- W^-)$  & 110 & 140 & 140 & 170 \\
$B$(background) & 710 & 710 & 710 & 710\\
$S/\sqrt B$ & 4.0 & 5.2 & 5.4 & 6.3 \\
\hline
\end{tabular}
\end{center}
\label{eeWW}
\end{table}

\begin{figure}
\centerline{\psfig{file=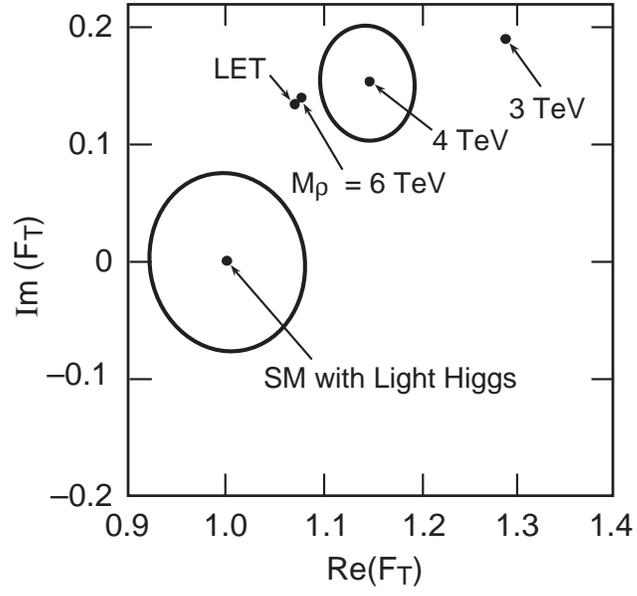,width=0.7\textwidth}}
\caption[technirho]{Confidence level contours for the real and imaginary
parts of 
$F_T$ at $\protect \sqrt{s} = 1.5$~TeV with 190 fb$^{-1}$.
The initial state electron polarization is 90\%. The contour about the
light Higgs boson value of $F_T=(1,0)$ is 95\% confidence level and
the contour about the $M_\rho=4$~TeV point is 68\% confidence level.}
\label{contours}
\end{figure}

If there is a techni-$\rho$ resonance, an electron positron collider
will have an ideal signal.  The production of $W$-pairs has one of the
biggest cross sections at a future $e^+ e^-$ linear collider.  If the
$W$-bosons in the final state are longitudinally polarized, they can
rescatter due to a tail of the techni-$\rho$ resonance.  The
rescattering modifies various final state distributions of the $W$ decay
products.  Studies show
that one can see the effects of a techni-$\rho$ up to 2~TeV at 95~\%
confidence level at a linear collider with $\sqrt{s} = 500$~GeV and
50~fb$^{-1}$,\cite{HIM} which is already comparable to the reach at the
LHC.  
Fig.~\ref{contours} contains confidence level contours for the real and
imaginary parts ofthe rescattering amplitude $F_T$ at $\sqrt{s}=1.5$ TeV
with 190 fb$^{-1}$.\cite{Barklow}
Shown are the 95\% confidence level contour about the
light Higgs boson value of $F_T$, as well as the 68\% confidence level
contour about the value of $F_T$
for a 4 TeV techni-$\rho$.
Even the non-resonant LET point is well outside the light Higgs boson
95\% confidence level region.
The  6~TeV and
and 4~TeV techni-$\rho$ points correspond to 4.8$\sigma$ and
6.5$\sigma$ signals, respectively.
At a slightly higher integrated luminosity
of 225 fb$^{-1}$, it is possible to obtain 7.1$\sigma$, 5.3$\sigma$ and
5.0$\sigma$ signals for
a 4~TeV techni-rho, a 6~TeV techni-rho, and LET, respectively.

The signatures of strong electroweak symmetry breaking sector discussed
so far are $WW$ scattering and are relatively model-independent.
There are signatures relevant at lower energies, though more
model-dependent.  Since our aim is to sort out the correct model which
describes the electroweak symmetry breaking, such model-dependence is
of great interest.  Now we turn our discussions to the model-dependent
signatures. 

First of all, one needs to recall that the scenario of strongly
interacting electroweak symmetry breaking sector has many problems.  Just
to name a few, Peskin--Takeuchi $S$-parameter, flavor-changing
neutral currents, typically too small $m_t$, large isospin splitting
$m_b \ll m_t$, $R_b$, etc.  Since it is not so useful to discuss
experimental signatures of models which are already excluded, I would
like to discuss several attempts to cure some of the above problems.
Interestingly enough, such attempts tend to give us signatures at lower
energies than a model-independent discusssion gives.

The first example is the $Z t\bar{t}$ vertex.  Suppose technicolor
theory is right, in the sense that the source of $W$, $Z$ and all
fermion masses originate from a single technifermion condensate $\langle
\overline{T} T \rangle$.  Since $m_t$ is large, $\simeq 175$~GeV, there needs
to be a fairly strong four-fermi interaction, $\bar{t} t \overline{T} T$.
Such an operator can be generated by an exchange of Extended Technicolor
(ETC) gauge boson which converts a standard model fermion (top quark in
this case) to a techni-fermion.  Exchange of such an ETC gauge boson
gives an interesting contribution to the $Z b \bar{b}$ vertex.\cite{CSS}
The naive ETC model reduces $R_b$, which is the wrong
direction given the current tendency in experimental data.  There are
two modified ETC models which give positive contributions to $R_b$, a
diagonal ETC boson\cite{Kitazawa} and a non-commuting ETC gauge
boson.\cite{CST}   In each case, one can choose a parameter such that the 
additonal contribution is consistent with the current value of $R_b$.
Interesting point is that these models tend to give a rather large
correction to $Z t\bar{t}$ vertex.  An analysis\cite{SB} shows 
one can measure vector and axial form factors of the top quark at 10~\%
level with 50~fb$^{-1}$ for each electron beam polarization at $\sqrt{s}
= 400$~GeV.  The predicted values of the vector form factor falls
typically outside the 95~\% confidence level contour.  

\begin{figure}
\centerline{\psfig{file=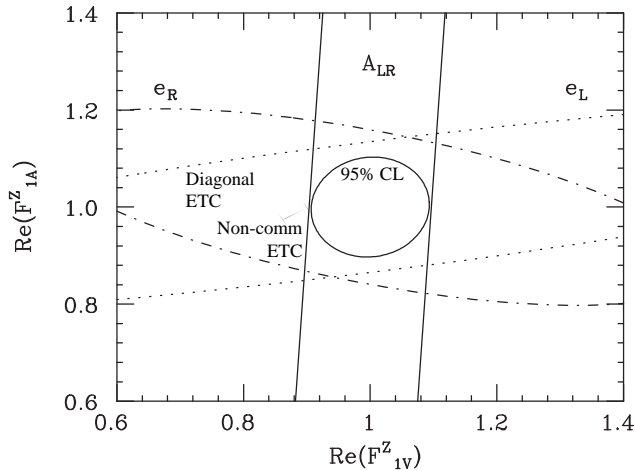,width=0.7\textwidth}}
\caption[FAFV]{95\% confidence level contours
for $F_{1V}^Z$ and $F_{1A}^Z$, obtained from the maximum-likelihood
analysis using a sample of 50 fb$^{-1}$ each of right- and
left-polarized electrons.\cite{SB}  Predictions from
diagonal ETC\cite{Kitazawa} and non-commuting ETC\cite{CST} put
in.\cite{preparation}}  
\label{FAFV}
\end{figure}

Another interesting model is an attempt to reduce the $S$-parameter
which tends to be too large.  Since the minimal model of technicolor,
one-doublet model with $N_{TC}=2$, is now excluded at more than 99~\%
confidence level,\cite{Hagiwara-Beijing} one needs to find a mechanism
to reduce the $S$-parameter.  An attempt by Appelquist and
Terning\cite{AT} is to 
introduce large isospin splitting, thereby sacrificing $T$-parameter a
little, to reduce the $S$-parameter even in a one-family model.  Their
point is that one can have techni-leptons to be rather light; then the
contribution to $T$-parameter can be small enough even when there is a
large isospin splitting between techni-electron and techni-neutrino.
Their sample spectrum of techni-fermions is
\par
\begin{center}
\begin{tabular}{ccrl}
$N$ & & 50 & GeV,\\
$E$ & & 150 & GeV,\\
$Q$ & & 600 & GeV.
\end{tabular} 
\end{center}
\par
\noindent
As apparent from the spectrum, this model predicts light techni-$\rho$,
$\overline{N} N$ at 100--300~GeV and a light charged
pseudo-Nambu-Goldston boson $N\overline{E}$ at 50--150~GeV.  This
techni-$\rho$ does not contribute much to the $WW$ rescattering because
the techni-neutrino contributes little to the $W$ and $Z$ masses.
However it can appear as a narrow resonance in $e^+ e^-$ collision.
$N\bar{E}$ can be produced similar to a charged Higgs boson whose main
decay mode is $\nu_\tau \tau^+$.  A search for it is straight forward,
looking for acoplanar $\tau$-pairs using right-handed electron
polarization to suppress the $WW$ background.  On the other hand there
are many colored psudo-Nambu-Goldstone bosons at 250--500~GeV, which are
targets of experiments at the LHC.

There are also attempts to solve the problem of flavor-changing neutral
currents which have typically too large rates in extended technicolor
models.  The 
mechanism called techni-GIM\cite{techni-GIM} is one of such attempts.
It requires a very 
complicated gauge strcture and needs many new fermion fields below 1~TeV
to cancel anomalies.  There typically are many pseudo-Nambu-Goldstone
bosons as well, whose masses arise due to gauge interactions.  Since
colorless pseudo-Nambu-Goldstone bosons are typically lighter than
colored one, the situation is quite similar to the supersymmetry.  The
LHC will look for colored ones, a linear collider for colorless ones.

Summarizing this section, a combination of two observations, (1) the
absolute absense of a Higgs boson at a linear collider and (2) a slight
excess in $WW$-scattering at the LHC can be a convincing signature of
strong electroweak sector.  Moreover, the excess in $WW$-scattering
observed at the LHC can be cross-checked with a linear collider at
$\sqrt{s} = 1.5$~TeV; if the excess is due to a techni-$\rho$, $\sqrt{s}
= 500$~GeV may be already enough.  There are other model-dependent
signatures like $Zt\bar{t}$ coupling, pseudo-Nambu-Goldstone bosons,
light techni-resonances, etc, which help sorting out the correct model.
Here again it is clear that the combination of both types of colliders
is important to understand physics of electroweak symmetry breaking.

\section{Conclusion}

Particle Physics is alive and well, it is approaching the most exciting
stage of experiments exploring the physics of electroweak symmetry
breaking.  The combination of the LHC and an electron positron linear
collider will allow us to sort out scenarios of electroweak
symmetry breaking.  Having only one of them may lead to an ambiguous and
unsatisfactory exploration of the physics, while having both can
give us hints to physics at yet higher energy scales.

\section*{Acknowledgments}

I express sincere thanks to the organizers of this workshop.  
I also thank Jonathan L. Feng for his useful comments on the draft.
This work was supported in part by the Director, Office of
Energy Research, Office of High Energy and Nuclear Physics, Division of
High Energy Physics of the U.S. Department of Energy under Contract
DE-AC03-76SF00098 and in part by the National Science Foundation under
grant PHY-90-21139.

\end{document}